\documentclass[conference]{IEEEtran}
\usepackage{amssymb}
\usepackage{cite}
\ifCLASSINFOpdf
  \usepackage[pdftex]{graphicx}
\else
  \usepackage[dvips]{graphicx}
\fi
\usepackage{algorithmic}
\usepackage{array}
\ifCLASSOPTIONcompsoc
  \usepackage[caption=false,font=normalsize,labelfont=sf,textfont=sf]{subfig}
\else
  \usepackage[caption=false,font=footnotesize]{subfig}
\fi
\usepackage{url}
\usepackage{color}
\usepackage{multirow}
\usepackage{multicol}
\usepackage{amsmath,graphicx}
\usepackage{pgfplots}
\usepackage{xspace}
\usepackage{siunitx}
\usepackage{bm}
\usepackage{listings}
\usepackage{tkz-kiviat}
\usepackage{array}
\usepackage{tabularx}
\usepackage{parskip}
\usepackage[vlined, linesnumbered, ruled]{algorithm2e}
\usepackage{flushend}
\usetikzlibrary{arrows}

\newcommand{\Leonardon}{L\'{e}onardon}
\newcommand{\Jego}{J\'{e}go}
\newcommand{\Montreal}{Montr\'{e}al}
\newcommand{\Ecole}{\'{E}cole }

\newcommand{\specialcell}[2][c]{%
  \begin{tabular}[#1]{c}#2\end{tabular}}
\newcommand{\snrcell}{\specialcell{$\bm{E_b/N_0}$\\\textbf{loss}\\\textit{dB}}}
\newcommand{\thrcell}{\specialcell{\textbf{Thr.}\\\textbf{gain}\\\textit{\%}}}

\newcommand{\matrixfootnote}{\footnote{$F^{\otimes 1} = 
  \begin{bmatrix}
  1 & 0 \\
  1 & 1
  \end{bmatrix}
  $ and $\forall n > 1, 
  F^{\otimes n} = 
  \begin{bmatrix}
  F^{\otimes n - 1} & 0_{n-1} \\
  F^{\otimes n - 1} & F^{\otimes n - 1}
  \end{bmatrix}$, where $n=\log_2(N)$, $N$ is the codeword length, and $0_{n}$ is a $2^n$-by-$2^n$ matrix of zeros.
}}
\begin{document}

\title{Fast and Flexible Software Polar List Decoders}
% TODO Discuss about title

\author{\IEEEauthorblockN{Mathieu \Leonardon\IEEEauthorrefmark{1}\IEEEauthorrefmark{2},
Adrien Cassagne\IEEEauthorrefmark{1},
Camille Leroux\IEEEauthorrefmark{1},
Christophe \Jego\IEEEauthorrefmark{1},\\
Louis-Philippe Hamelin\IEEEauthorrefmark{3} and
Yvon Savaria\IEEEauthorrefmark{2}
}
\IEEEauthorblockA{\IEEEauthorrefmark{1}IMS Laboratory, UMR CNRS 5218, Bordeaux INP, University of Bordeaux, Talence, France}
\IEEEauthorblockA{\IEEEauthorrefmark{2}\Ecole Polytechnique de \Montreal, QC, Canada}
\IEEEauthorblockA{\IEEEauthorrefmark{3}Huawei Technologies Canada Co. LTD, Ottawa, ON, Canada}} 

\maketitle

\begin{abstract}
 Flexibility is one mandatory aspect of channel coding in modern wireless communication systems. Among other things, the channel decoder has to support several code lengths and code rates. This need for flexibility applies to polar codes that are considered for control channels in the future 5G standard. This paper presents a new generic and flexible implementation of a software Successive Cancellation List (SCL) decoder. A large set of parameters can be fine-tuned dynamically without re-compiling the software source code: the code length, the code rate, the frozen bits set, the puncturing patterns, the cyclic redundancy check, the list size, the type of decoding algorithm, the tree-pruning strategy and the data quantization. This generic and flexible SCL decoder enables to explore tradeoffs between throughput, latency and decoding performance. Several optimizations are proposed to achieve a competitive decoding speed despite the constraints induced by the genericity and the flexibility. The resulting polar list decoder is about 4 times faster than a generic software decoder and only 2 times slower than a non-flexible unrolled decoder. Thanks to the flexibility of the decoder, the fully adaptive SCL algorithm can be easily implemented and achieves higher throughput than any other similar decoder in the literature (up to 425 Mb/s on a single processor core for N = 2048 and K = 1723 at 4.5 dB).
\end{abstract}


\IEEEpeerreviewmaketitle

\section{Introduction}
\label{sec:intro}

  Polar codes \cite{arikan09} are the first provably capacity achieving channel codes, for an infinite code length. The decoding performance of the original Successive Cancellation (SC) decoding algorithm is however not satisfactory for short polar codes. The Successive Cancellation List (SCL) decoding algorithm has been proposed in \cite{tal12} to counter this fact along with the concatenation of a Cyclic Redundancy Check (CRC). The decoding performance of SCL decoding is such that polar codes is included in the fifth generation (5G) mobile communications standard \cite{3GPP_16}.
  
  Cloud radio access network (Cloud-RAN) is foreseen by both academic \cite{wubben2014benefits,rost2014cloud} and industrial \cite{ericsson-wp-cloud-ran,huawei-5G} actors as one of the key technologies of the 5G standard. In the Cloud-RAN the virtualization of the physical layer (PHY) would allow for deep cooperative multipoint processing and computational diversity \cite{wubben2014benefits}. PHY-layer cooperation enables interference mitigation, while computational diversity lets the network balance the computational load accross multiple users. But the virtualization of the FEC decoder is a challenge as it is one of the most computationally intensive tasks of the signal processing chain in a Cloud-RAN context\cite{rodriguez2017towards,nikaein2015processing}. Therefore, efficient, flexible and parallel software implementations of FEC decoders are needed to enable some of the expected features of Cloud-RAN.

  To date, the fastest software implementations of SCL polar decoders have been proposed in \cite{sarkis16}. The high decoding speed is achieved at the price of flexibility, because the software decoder is only dedicated to a specific polar code. In a wireless communication context, the source code of this fast software polar decoder would have to be recompiled every time the Modulation and Coding Scheme (MCS) changes, which may happen every millisecond.

  In this work, we propose a software SCL polar decoder able to switch between different channel coding contexts (block length, code rate, frozen bits sets, puncturing patterns and CRC code). This property is denoted as \textit{genericity}. Moreover, the proposed decoder supports different list-based decoding algorithms, several list sizes ($L$), quantization formats and tree-pruning techniques during a real time execution. Again, this is done dynamically without having to recompile the software description. We denote this feature as \textit{flexibility}. The genericity and the flexibility of the decoder are achieved without sacrificing the decoding throughput and latency thanks to several implementation optimizations. Actually, the proposed software SCL decoder is only 2 times slower than a polar code specific decoder \cite{sarkis16} and 4 times faster than a generic decoder \cite{sarkis14_3}. Unlike these fast decoders, the proposed decoder supports a fully adaptive version of SCL. It reaches 425 Mb/s on a single processor core for $N = 2048$ and $K = 1723$ at 4.5 dB.

  The genericity of our decoder makes it compliant with a wireless communication context: one can change the polar code parameters dynamically. Thanks to the decoder flexibility, some new tradeoffs between throughput and error rate performance are now possible. Finally, besides the genericity/flexibility-driven improvements, some specific optimizations were implemented in order to match the state-of-the-art throughputs of software SCL decoders. Among other optimizations, a new sorting technique is applied to different parts of the algorithm which is faster than any other according to our experimentations. New methods to speed up the CRC processing are also applied. The polar functions library described in \cite{cassagne15,cassagne16_2} is used in order to benefit from a portable implementation of SIMD instructions.
     
  The rest of the paper is organized as follows: Section \ref{sec:polar_codes} describes the SCL decoding algorithm and the improved versions. The genericity and the flexibility of the proposed decoder are highlighted in Section \ref{sec:genericity}. Section \ref{sec:implem_improv} details the speed-oriented optimizations. Finally, Section \ref{sec:measures} provides the throughput and latency performance.

\section{Polar Codes}
\label{sec:polar_codes}

  \begin{figure}[t]
  \centering
  \includegraphics[width=0.49\textwidth]{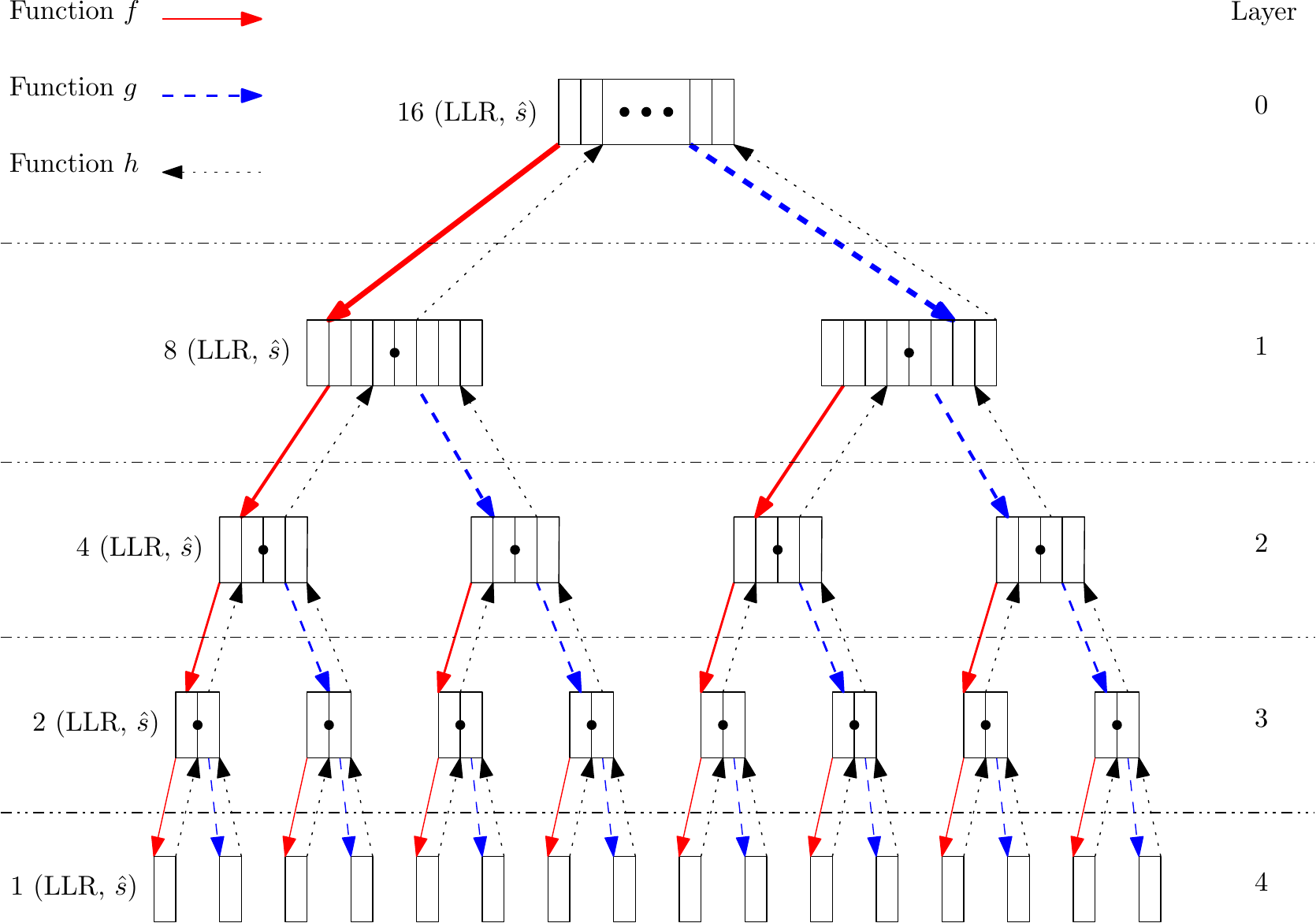}
  \caption{Full SC decoding tree ($N = 16$).}
  \label{fig:dec_tree}
  \end{figure}
  
  In this section, we first present the polar encoding process. Then the SC and SC-List based decoding algorithms are reviewed. Finally we discuss the tradeoffs between speed and decoding performance of different decoding algorithms.
      
\subsection{Polar Encoding Process}

  In the polar encoding process, an information sequence $\bm{b}$ of length $K$ is transformed into a codeword $\bm{x}$ of length $N$. The first step is to build a vector $\bm{u}$ in which the information bits $\bm{b}$ are mapped on a subset $\bm{u}_{\mathcal{A}}$ where $\mathcal{A}\subset\{0,...,N-1\}$. The remaining bits $\bm{u}_{\mathcal{A}^c} = (a_i : i\not\in\mathcal{A})$ called \textit{frozen bits} are usually set to zero. The selection of the frozen bits is critical for the effectiveness of the polar codes. Two of the main techniques to date for constructing polar codes are based on the Density Evolution approach \cite{6557004} and on the Gaussian Approximation \cite{6279525}. These techniques sort the polar channels according to their reliability in order to choose the frozen bits set for a given code length. Then, an intermediate vector $\bm{u'}$ is generated thanks to an encoding matrix\matrixfootnote: $\bm{u'} = \bm{u}F^{\otimes n}$. Finally the bits in the subset $\bm{u'_}{\mathcal{A}^c}$ are set to zero and the output codeword is $\bm{x} = \bm{u'}F^{\otimes n}$. This encoding method is called systematic because the \textit{information sequence} $\bm{b}$ is present in the codeword ($\bm{x}_{\mathcal{A}}=\bm{b}$). In this paper, systematic encoding schemes are considered. A CRC of length $c$ may be concatenated to the information sequence $\bm{b}$ in order to improve the decoding performance of SCL decoding algorithms. In this case, $|\mathcal{A}|=K+c$ and the CRC is included in $\bm{u}_{\mathcal{A}}$. In this paper, the code rate is defined as $R=K/N$ and the $c$ bits of the CRC are not considered as information bits. For instance, a polar code whose block length is $N=2048$ and code rate is $R = 1/2$ contains 1024 informations bits. Such a code is denoted as ($2048$,$1024$).

  \subsection{Polar Decoding Algorithms}

  \subsubsection{SC decoding algorithm}
  
  The SC decoding process can be seen as the pre-order traversal of a binary tree as shown in Figure~\ref{fig:dec_tree}. The tree contains $\log_2 N + 1$ layers. Each layer  
  contains $2^d$ nodes, where $d$ is the depth of the layer in the tree. Each node contains a set of $2^{n-d}$ Log-Likelihood Ratios (LLRs) $\lambda$ and partial sums $\hat{s}$. The partial sums correspond to the propagation towards the top of the tree of hard decisions made in the \textit{update\_paths()} function. As shown in Figure~\ref{fig:dec_tree}, LLRs, which take real values, and partial sums, which take binary values, are the two types of data contained in the decoding tree, and three functions, $f$, $g$ and $h$ are necessary for updating the nodes:
  {
    \begin{eqnarray*}
      \left\{\begin{array}{l c l}
        f(\lambda_a,\lambda_b) &=& sign(\lambda_a.\lambda_b).\min(|\lambda_a|,|\lambda_b|)\\
        g(\lambda_a,\lambda_b,\hat{s}_a)&=&(1-2\hat{s}_a)\lambda_a+\lambda_b\\
        h(\hat{s}_a,\hat{s}_b)&=& (\hat{s}_{a} \oplus \hat{s}_{b}, \hat{s}_{b})
      \end{array}\right.
      \label{eq:f_g}
    \end{eqnarray*}
  }
  In comparison with the SCL algorithm and its derivatives, the computational complexity of the SC algorithm is low: $O(N\log_2N)$. Therefore, both software \cite{legal15} and hardware \cite{sarkis14_1} implementations achieve multi-Gb/s throughputs with low latencies. The drawback of the SC decoding algorithm is its decoding performance especially for short polar codes. This is an issue for the future 5G wireless standard in which polar codes are targeted for control channels, with code lengths shorter than $2048$ \cite{3GPP_16}.
  
  \subsubsection{SCL decoding algorithm}
  
    \begin{algorithm}%[t]
    \label{alg:scl}
    
    \small
    \SetKwProg{Fn}{Function}{}{}
    
    % \KwIn{$N$ is the frame size.}
    % \KwIn{$L$ is the number of lists (or paths) to maintain.}
    \KwData{$\lambda$ is a 2D buffer ($[L][2N]$) to store the LLRs.}
    \KwData{$\hat{s}$ is a 2D buffer ($[L][N]$) to store the bits.}
    
    \Fn{SCL\_decode ($N, o_{\lambda}, o_{\hat{s}}$)}
    {
      $N_{\frac{1}{2}} = N / 2$
      
      \uIf(// not a leaf node){$N > 1$}
      {
        \For(// loop over the paths){$p=0$ \textbf{to} $L-1$}
        {
          \For(// apply the $f$ function){$i=0$ \textbf{to} $N_{\frac{1}{2}}-1$}
          {
            $\lambda[p][o_\lambda + N + i] = \bm{f}(\lambda[p][o_\lambda + i], \lambda[p][o_\lambda + N_{\frac{1}{2}} + i])$
          }
        }
        
        \textit{SCL\_decode ($N_{\frac{1}{2}}, o_{\lambda} + N, o_{\hat{s}}$)}
        
        \For{$p=0$ \textbf{to} $L-1$}
        {
          \For(// apply the $g$ function){$i=0$ \textbf{to} $N_{\frac{1}{2}}-1$}
          {
            $\lambda[p][o_\lambda + N + i] = \bm{g}(\lambda[p][o_\lambda + i], \lambda[p][o_\lambda + N_{\frac{1}{2}} + i], \hat{s}[p][o_{\hat{s}} + i])$
          }
        }
        
        \textit{SCL\_decode ($N_{\frac{1}{2}}, o_{\lambda} + N, o_{\hat{s}} + N_{\frac{1}{2}}$)}
        
        \For{$p=0$ \textbf{to} $L-1$}
        {
          \For(// update the partial sums){$i=0$ \textbf{to} $N_{\frac{1}{2}}-1$}
          {
            $\hat{s}[p][o_{\hat{s}} + i] = \bm{h}(\hat{s}[p][o_{\hat{s}} + i], \hat{s}[p][o_{\hat{s}} + N_{\frac{1}{2}} + i])$
          }
        }
      }
      \Else(// a leaf node)
      {
        \textit{update\_paths ()} // update, create and delete paths
      }
    }
    
    \textit{SCL\_decode ($N, 0, 0$)} // launch the decoder
    
    \textit{select\_best\_path ()}
    
    \caption{SCL decoding algorithm}
  \end{algorithm}

   The SCL algorithm is summarized in Algorithm~\ref{alg:scl}. Unlike the SC algorithm, the SCL decoder builds a list of candidate codewords along the decoding. At each call of the \textit{update\_paths()} sub-routine (Alg.~\ref{alg:scl}, l.16), $2L$ candidates are generated. A path metric is then evaluated to keep only the $L$ best candidates among the $2L$ paths. The path metrics are calculated as in \cite{balatsoukas2015llr}. At the end of the decoding process, the candidate codeword with the best path metric is selected in the \textit{select\_best\_path()} sub-routine (Alg.~\ref{alg:scl}, l.18). 
   The decoding complexity of the SCL algorithm grows as $O(LN\log_2N)$. This linear increase in complexity with L leads to significant improvements in BER/FER performances, especially for small code lengths.
  
  \subsubsection{Simplified SC and SCL decoding algorithms}
  
    All aforementioned polar decoding algorithms have in common that they can be seen as a pre-order tree traversal algorithm. In \cite{alamdar-yazdi11}, a tree pruning technique called the Simplified SC (SSC) was applied to SC decoding. An improved version was proposed in \cite{sarkis14_1}. This technique relies on the fact that, depending on the frozen bits location in the leaves of the tree, the definition of dedicated nodes enables to prune the decoding tree: Rate-0 nodes (\texttt{R0}) correspond to a sub-tree whose all leaves are frozen bits, Rate-1 nodes (\texttt{R1}) correspond to a sub-tree in which all leaves are information bits, REPetition (\texttt{REP}) and Single Parity Check (\texttt{SPC}) nodes correspond to repetition and SPC codes sub-trees. These special nodes, originally defined for SC decoding, can be employed in the case of SCL decoding as long as some modifications are made in the path metric calculation \cite{sarkis16}. This tree-pruned version of the algorithm is called Simplified SCL (SSCL).
    The tree pruning technique can drastically reduce the amount of computation in the decoding process. Moreover, it increases the available parallelism by replacing small nodes by larger ones. As will be discussed in Section~\ref{sec:genericity}, the tree pruning may have a small impact on decoding performance.
    
  \subsubsection{CRC concatenation scheme}
  
  The authors in \cite{tal12} observed that when a decoding error occurs, the right codeword is often in the final list, but not with the best path metric. They proposed to concatenate a CRC to the codeword in order to discriminate the candidate codewords at the final stage of the SCL decoding. Indeed, this technique drastically improves the FER performance of the decoder. We denote this algorithm CA-SCL and its simplified version CA-SSCL. In terms of computational complexity, the overhead consists in the computation of $L$ CRC at the end of each decoding. 
    
  \subsubsection{Adaptive SCL decoding algorithm}

  The presence of the CRC can be further used to reduce the decoding time by gradually increasing $L$. This variation of SCL is called Adaptive SCL (A-SCL) \cite{li12}. The first step of the A-SCL algorithm is to decode the received frame with the SC algorithm. Then, the decoded polar codeword is checked with a CRC. If the CRC is not valid, the SCL algorithm is applied with $L=2$. If no candidate in the list satisfies the CRC, $L$ is gradually doubled until it reaches the value $L_{max}$. In this paper, we call this version of the A-SCL decoding the Fully Adaptive SCL (FA-SCL) as opposed to the Partially Adaptive SCL (PA-SCL), in which the $L$ value is not gradually doubled but directly increased from $1$ (SC) to $L_{max}$. The simplified versions of these algorithms are denoted PA-SSCL and FA-SSCL. In order to simplify the algorithmic range, in the remainder of the paper, only the simplified versions are considered. The use of either FA-SSCL or PA-SSCL algorithmic improvement introduces no BER or FER performance degradation as long as the CRC length is adapted to the polar code length. If the CRC length is too short, the decoding performance may be degraded because of false detections. These adaptive versions of SSCL can achieve higher throughputs. Indeed, a large proportion of frames can be decoded with a single SC decoding. This is especially true when the SNR is high. This will be further discussed in Section~\ref{sec:genericity}.

  % TODO : Christophe proposait de rajouter "an error floor as explained in...". Le problème c'est que ce n'est expliqué nulle part

  \begin{table}[t]
    \centering
    \caption{Throughput and latency comparison of polar decoding algorithms.}
    \label{tab:algo}
    {\small\resizebox{\linewidth}{!}{
     \begin{tabular}{r|c|c|c} 
      \textbf{Decoding}  & \textbf{BER \& FER}   & \multirow{1}{*}{\textbf{Throughput}} & \textbf{Max. Latency}         \\
      \textbf{Algorithm} & \textbf{Performances} & ($\bm{\mathcal{T}}$)                 & ($\bm{\mathcal{L}_{worst}}$)  \\
      \hline
      SC        & poor       & medium & medium \\
      SSC       & poor       & high   & low    \\
      SCL       & good       & low    & high   \\
      SSCL      & good       & low    & medium \\
      CA-SSCL   & very good  & low    & medium \\
      PA-SSCL   & very good  & high   & medium \\
      FA-SSCL   & very good  & high   & high   \\
    \end{tabular}
    }}
  \end{table}

  % Tree pruning is the key point to achieve high throughput and small latency for to main reasons: 1) it significantly decreases the complexity of the algorithm, 2) it reduces the sequentiality by replacing small nodes (see layers 4,3, etc. in Fig.~\ref{fig:dec_tree}) by bigger nodes where there is more parallelism.

\subsection{Algorithmic Comparison}

  In order to better distinguish all the algorithmic variations, we compare their main features in Table~\ref{tab:algo}. Each algorithm is characterized in terms of decoding performance, throughput, and worst case latency for a software implementation. The non-simplified versions of the adaptive SCL algorithms are not included in the Table for readability.
  
  The SC and especially the SSC algorithms achieve very high throughput and low latency with poor BER and FER performances. The SCL algorithm improves the decoding performance compared to the SC algorithm, but its computational complexity leads to an increased latency and a lower throughput. The SSCL algorithm improves the decoding throughput and latency without any impact in terms of BER and FER performances, as long as the tree pruning is not too deep, as will be discussed in Section~\ref{sec:genericity}. Therefore, tree pruning is applied to all the following algorithms, namely CA-SSCL, FA-SSCL and PA-SSCL. By applying CRC to the SCL algorithm, one can achieve better BER and FER performances at the cost of computational complexity overhead. The Adaptive SCL algorithms reduce the decoding time with no impact on BER and FER performances. Furthermore, a tradeoff between throughput and worst case latency is possible with the use of either PA-SSCL or FA-SSCL decoding algorithms.

  \begin{figure}[t]
  \centering
  \includegraphics[width=0.49\textwidth]{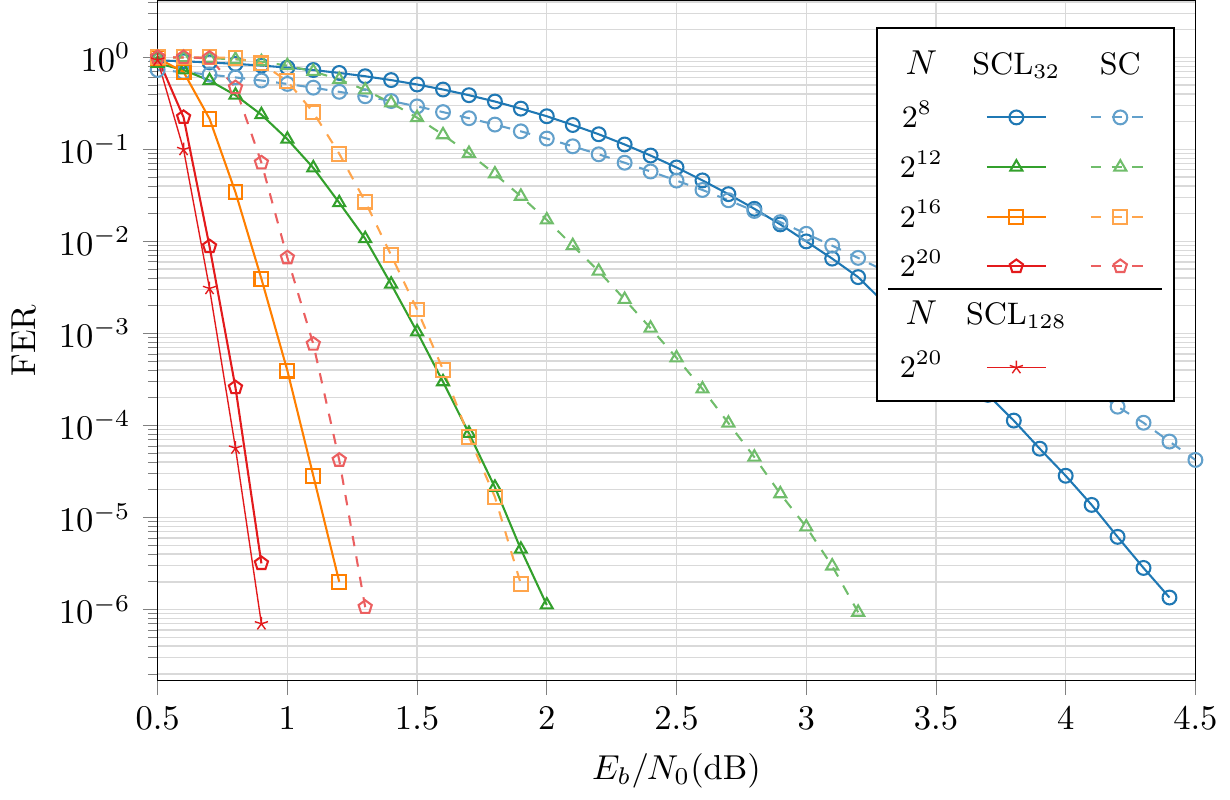}
  \caption{Decoding performance comparison between CA-SCL and SC decoders.
           Code rate $R = 1/2$, and 32-bit CRC (GZip).}
  \label{plot:code}
  \end{figure}

 To the best of our knowledge, SC and CA-SCL decoding performances have never been investigated in the literature for large code lengths ($N>2^{14}$). This is probably due to the long simulation durations. In this work the AFF3CT\footnote{AFF3CT is an Open-source software (MIT license) for fast forward error correction simulations, see \texttt{http://aff3ct.github.io}} tool enables multi-threaded and multi-nodes simulations.
 All the presented simulations use the Monte Carlo method with a Binary Phase-Shift Keying (BPSK) modulation. The communication channel is an Additive White Gaussian Noise (AWGN) channel based on the Mersenne Twister pseudo-random number generator (MT19937) \cite{matsumoto1998mersenne} and the Box-Muller transform \cite{box1958note}.
 Figure~\ref{plot:code} compares the BER/FER performances of CA-SCL with SC decoding for a large range of code lengths. As expected, it appears that the coding gain brought by the SCL algorithm decreases for larger $N$ values. In the case of $N=2^{16}$, the improvement caused by the use of the CA-SCL algorithm with $L=32$ and a 32-bit GZip CRC (\texttt{0x04C11DB7} polynomial) instead of SC is about $0.75$ dB compared to $1.2$ dB with a polar code of size $N=2^{12}$. For larger polar codes, $N=2^{20}$, the gain is reduced to $0.5$ dB, even with a list depth of $128$ that is very costly in terms of computational complexity.

The tradeoffs between speed and decoding performance show some general trends. However, the efficiency of each decoding algorithm is strongly dependent on the polar code length, code rate, list depth and code construction. It is expected that the best tradeoff is not always obtained with a single algorithm and parameter set combination. It is consequently highly relevant to use a generic and flexible decoder, that supports all variants of the decoding algorithms. Thus, it is possible to switch from one to another as shown in the following section.

\section{Generic and Flexible Polar Decoder}
\label{sec:genericity}
  % Technique presented in\cite{sarkis14_1,cassagne15,sarkis16}.

  The main contribution of this work lies in the flexibility and the genericity of the proposed software decoder. These terms need to be clearly defined in order to circumvent possible ambiguity. In the remainder of the paper, the \textit{genericity} of the decoder concerns all the parameters that define the supported polar code such as the codeword length, the code rate, the frozen bits set, the puncturing patterns and the concatenated CRC. These parameters are imposed by the telecommunication standard or the communication context. In the wireless communications context, these are constantly adapted by AMC methods \cite{dahlman20134g}. In this work, a decoder is considered \textit{generic} if it is able to support any combination of these parameters that can be changed during a real time execution.
  On the other hand, the \textit{flexibility} of a decoder includes all the customizations that can be applied to the decoding algorithm for a given polar code: variant of the decoding algorithm, data representation format, list size $L$, tree pruning strategy, ... These customizations are not enforced by a standard. The flexibility gives some degrees of freedom to the decoder in order to find the best tradeoff between decoding performance, throughput or latency for a given polar code.
    
  \subsection{Genericity}
    In the context of wireless communications, the standards enforce several different code lengths $N$ that have to be supported to share bandwidth between different users. This is also the case for the code rate $R$ that needs to be adapted to the quality of the transmission channel. Therefore, a practical implementation should be adapted to both $N$ and $R$ in real-time in order to limit latency.
    
    A polar code is completely defined by $N$ and the frozen bits set $\bm{u}_{\mathcal{A}^c}$. Several methods exist to generate some "good" sets of frozen bits \cite{6557004,6279525}. The code rate $R$ depends on the size of $\bm{u}_{\mathcal{A}^c}$. In their original form, polar code lengths are only powers of two. The puncturing and shortening techniques in \cite{6936302,6655078,7152894} enable to construct polar codes of any length at the cost of slightly degraded decoding performance. The coding scheme can be completed with the specification of a CRC.

  In \cite{sarkis16}, the unrolling method is used: a specific description of the decoder has to be generated for a specific polar code parameter set of $N$, $K$, $R$, frozen bits set, puncturing pattern, CRC. This approach leads to very fast software decoders at the price of the genericity, since a new source code should be generated and compiled every time the modulation and coding scheme (MCS) changes. This method is not adapted to wireless communication standards, in which these parameters have to be adapted not only over time, but also for the different users.

    The proposed decoder does not use the unrolling method and is completely generic regarding the code dimension $K$, the code length $N$, the frozen bits set $\bm{u}_{\mathcal{A}^c}$ and the puncturing patterns. All of them are dynamic parameters of the decoder and can be defined in input files. All CRC listed in \cite{CRCWiki} are available along with the possibility to define others. It is shown in \cite{zhang2017crc} that custom CRCs for polar codes can have a very good impact on the decoding performance.

     Relying on an unique software description also implies that the tree pruning technique also has to be dynamically defined. Indeed, this technique depends on the frozen bits set $\bm{u}_{\mathcal{A}^c}$. Not sacrificing throughput or latency while maintaining the genericity imposed by wireless communication standards is at the core of the proposed implementation. Flexibility in terms of decoding algorithms, described in the following, along with improvements presented in Section~\ref{sec:implem_improv}, is necessary to deal with this challenge.

  \subsection{Flexibility}

  \begin{figure}[t]
  \centering
  \includegraphics[width=0.49\textwidth]{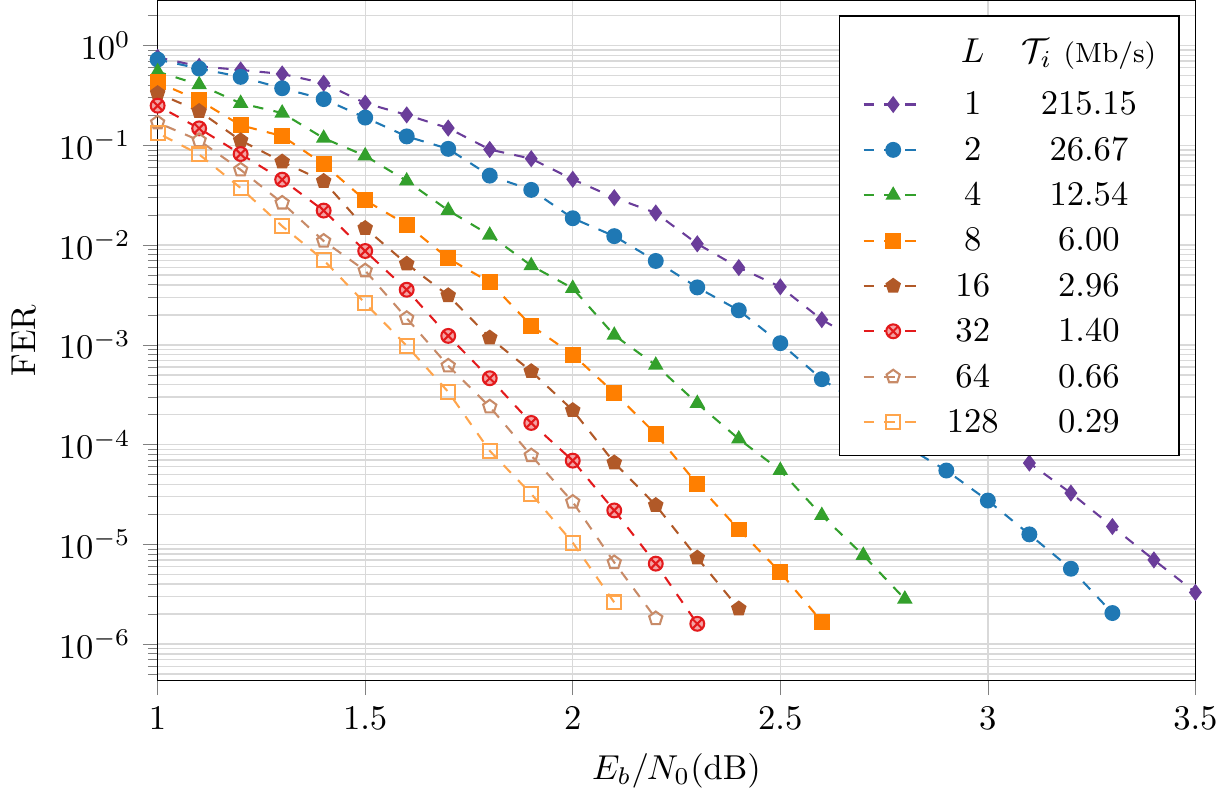}
  \caption{Tradeoffs between CA-SSCL decoding and throughput performances depending on $L$.
    $N=2048$, $R=0.5$, and 32-bit CRC (GZip).
    For $L=1$, the SSC decoder is used with a ($2048$,$1024$) polar code.}
  \label{plot:L}
  \end{figure}

  On one hand, the reason for the decoder genericity is the compliance to the telecommunication standards. On the other hand, the flexibility of the decoder regroups several algorithmic variations that are discussed in the following. These variations allow several tradeoffs of multiple sorts, whatever the standard. They are all included in a single source code.

  In the proposed decoders the following parameters can be changed dynamically without re-compilation: the list size $L$, the tree pruning strategy, the quantization of the LLRs and the different SCL variants. Each of these adjustments can be applied to access to different tradeoffs between throughput, latency, and error rate performance. As a consequence, one can easily fine-tune the configuration of the software decoder for any given polar code.

  \subsubsection{List size}
  
  As mentioned earlier, the list size $L$ impacts both speed and decoding performance. In Figure~\ref{plot:L}, the throughput as well as BER and FER performances of the CA-SSCL algorithm are shown for different $L$ values. A ($2048$,$1024$) polar code with a 32-bit CRC is considered. The computational complexity increases linearly with $L$: the throughput is approximately halved when $L$ is doubled, except for the case of the SC algorithm ($L=1$) which is much faster. Indeed, there is no overhead due to the management of different candidate paths during the decoding. For $L\geq4$ and $E_b/N_0=2$, the FER is also approximately halved when the list size $L$ is doubled.

  \subsubsection{Tree pruning strategy}

  \begin{figure}[t]
  \centering
  \includegraphics[width=0.48\textwidth]{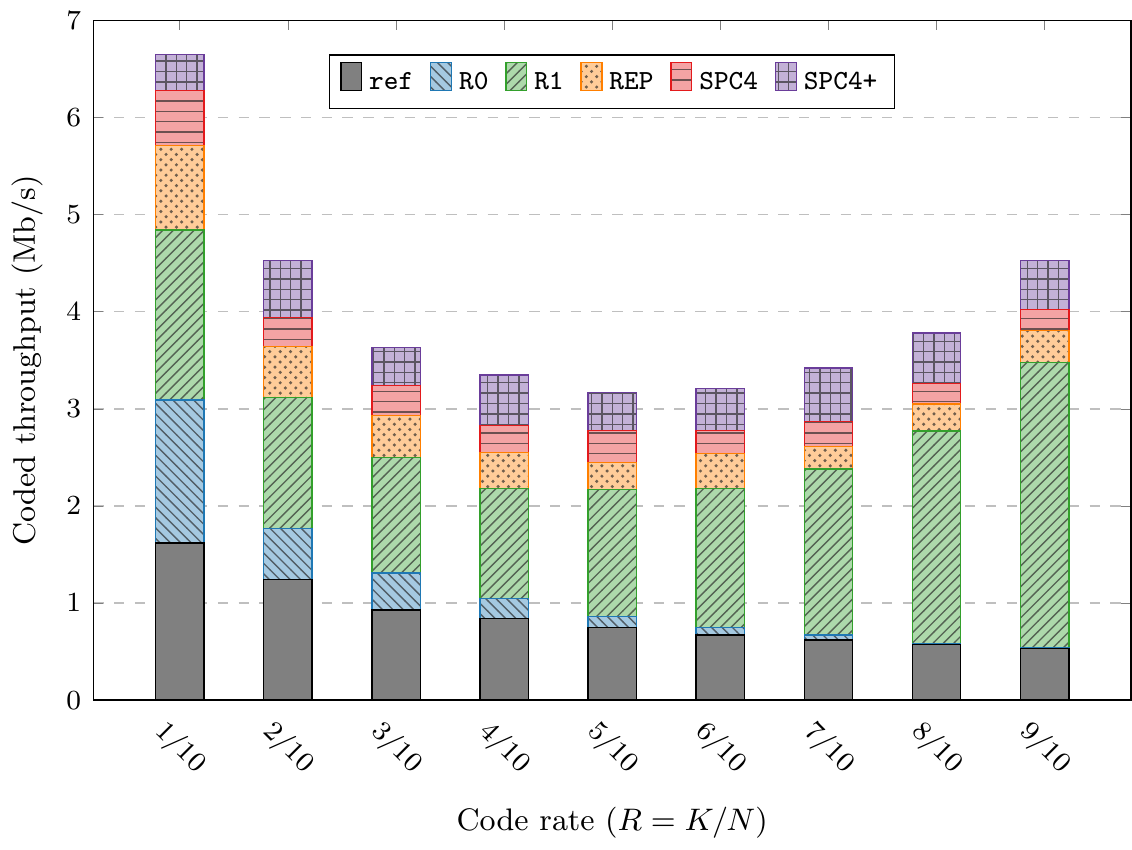}
  % \caption{Impact of the specialized nodes on the SCL coded throughput.
  \caption{Dedicated nodes impact on CA-SSCL. % Adrien j'ai racourci le titre pour des raisons de cosmétique
    $N=2048$ and $L=32$.}
  \label{plot:tree_cut}
  \end{figure}

  A second degree of flexibility is the customization of the SCL tree pruning. The authors in \cite{alamdar-yazdi11,sarkis16} defined dedicated nodes to prune the decoding tree and therefore to reduce the computational complexity. In this proposed decoder, each dedicated node can be activated separately. The ability to activate dedicated nodes at will is useful in order to explore the contribution of each node type on the throughput. Figure~\ref{plot:tree_cut} shows the impact of the different tree pruning optimizations on the CA-SSCL decoder throughput depending on the code rate. The performance improvements are cumulative. Coded throughput, in which the redundant bits are taken in account, is shown instead of information throughput, for which only information bits are considered in order to illustrate the computational effort without the influence of the fact that higher rate codes involve higher information throughput. 

  Without pruning, the coded throughput decreases as the code rate increases. Indeed, frozen bit leaf nodes are faster to process than information bit leaf nodes, in which a threshold detection is necessary. As there are more \texttt{R0} and \texttt{REP} nodes in low code rates, the tree pruning is more efficient in the case of low code rates. The same explanation can be given for \texttt{R1} nodes in high code rates. \texttt{R1} node pruning is more efficient than \texttt{R0} node pruning on average. Indeed, a higher amount of computations is saved in \texttt{R1} nodes than in \texttt{R0} nodes.

  It has also been observed in \cite{sarkis16} that when the \texttt{SPC} node size is not limited to $4$, the decoding performance may be degraded. Consequently the size is limited to $4$ in \texttt{SPC4}. In \texttt{SPC4+} nodes, there is no size limit. The two node types are considered in Figure~\ref{plot:tree_cut}. Therefore, the depth at which dedicated nodes are activated in the proposed decoder can be adjusted, in order to offer a tradeoff between throughput and decoding performance.

  \begin{table}[b]
  \centering
  \caption{Effects of the \texttt{SPC4+} nodes on the CA-SSCL @ $10^{-5}$ FER}
  \label{tab:spc4}
  {\small\resizebox{\linewidth}{!}{ 
\begin{tabular}{c|c|@{}c@{}|@{}c@{}|@{}c@{}|@{}c@{}|@{}c@{}|@{}c@{}}
  \multicolumn{2}{c|}{}  & \multicolumn{2}{c|}{$\bm{N = 256}$} & \multicolumn{2}{c|}{$\bm{N = 1024}$} & \multicolumn{2}{c}{$\bm{N = 4096}$} \\
  \hline
  \hline
  $\bm{L}$ & $\bm{R}$ & \snrcell & \thrcell & \snrcell & \thrcell & \snrcell & \thrcell \\
  \hline
  \hline
  \multirow{3}{*}{8}   &  1/3  &  0.15  &  09.7  &  0.03  &  12.6  &  0.02  &  09.5 \\
  \cline{2-8}          &  1/2  &  0.09  &  08.6  &  0.04  &  16.4  &  0.07  &  20.2 \\
  \cline{2-8}          &  2/3  &  0.03  &  20.5  &  0.04  &  11.3  &  0.09  &  14.3 \\
  \hline
  \hline 
  \multirow{3}{*}{32}  &  1/3  &  0.52  &  11.8  &  0.19  &  12.9  &  0.22  &  12.5 \\
  \cline{2-8}          &  1/2  &  0.30  &  10.3  &  0.24  &  16.5  &  0.26  &  19.9 \\
  \cline{2-8}          &  2/3  &  0.27  &  22.6  &  0.22  &  15.2  &  0.25  &  17.1 \\
\end{tabular}
}}
  \end{table}

  According to our experiments, the aforementioned statement about performance degradation caused by \texttt{SPC4+} nodes is not always accurate depending on the code and decoder parameters.
  The impact of switching \textit{on} or \textit{off} \texttt{SPC4+} nodes on decoding performance and throughput at a FER of $10^{-5}$ is detailed in Table~\ref{tab:spc4}. It shows that \texttt{SPC4+} nodes have only a small effect on the decoding performance. With $L=8$, an SNR degradation lower than 0.1 dB is observed, except for one particular configuration Throughput improvements of $8$ to $20$ percents are observed. If $L=32$, the SNR losses are more substantial (up to $0.5$ dB), whereas throughput improvements are approximately the same. Besides this observation, Table~\ref{tab:spc4} shows how the proposed decoder flexibility in the AFF3CT environment enables to optimize easily the decoder tree pruning, both for software implementations or for hardware implementations in which tree pruning can also be applied \cite{lin2014reduced}.

  \subsubsection{LLR Quantization}
  
  \begin{figure}[t]
  \centering
  \includegraphics[width=0.49\textwidth]{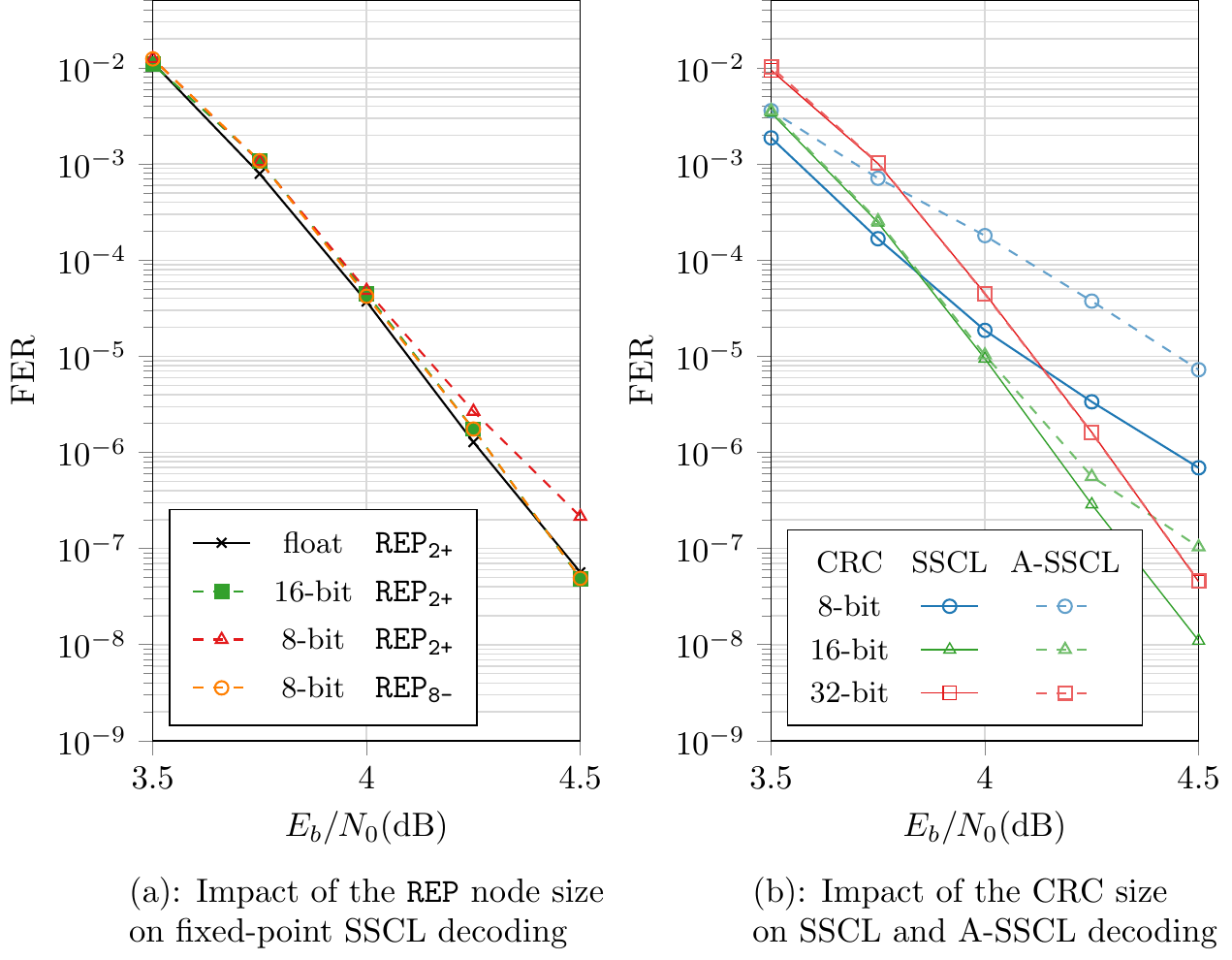}
  \caption{Decoding performance of the SSCL and the A-SSCL decoders. 
    Code ($2048$,$1723$), $L=32$.}
  \label{plot:ber_fer}
  \end{figure}

  \begin{table}[b]
    \centering
    \caption{Throughput and latency comparisons between floating-point (32-bit) and fixed-point (16-bit and 8-bit) Adaptive SSCL decoders. Code (2048,1723), $L = 32$ and 32-bit CRC (Gzip).}
    \label{tab:fixed}
    {\small\resizebox{\linewidth}{!}{ 
    \begin{tabular}{r | r | c || c | c || c | c || c | c}
      \multirow{2}{*}{\textbf{Decoder}} & \multirow{2}{*}{\textbf{Prec.}} & \multirow{2}{*}{$\bm{\mathcal{L}_{worst}}$} & \multicolumn{2}{c ||}{\textbf{3.5 dB}} & \multicolumn{2}{c ||}{\textbf{4.0 dB}} & \multicolumn{2}{c}{\textbf{4.5 dB}} \\
      \cline{4-9}
      & & & $\bm{\mathcal{L}_{avg}}$ & $\bm{\mathcal{T}_i}$ & $\bm{\mathcal{L}_{avg}}$ & $\bm{\mathcal{T}_i}$ & $\bm{\mathcal{L}_{avg}}$ & $\bm{\mathcal{T}_i}$ \\
      % \hline
      \hline
      \multirow{3}{*}{PA-SSCL} & 32-bit &  635 & 232.3 &   7.6 & 41.7 &  42.1 & 7.4 & 237.6 \\
      %\cline{3-9}
                               & 16-bit &  622 & 219.6 &   8.0 & 40.1 &  43.8 & 6.6 & 267.5 \\
      %\cline{3-9}
                               &  8-bit &  651 & 232.4 &   7.6 & 41.2 &  42.6 & 6.5 & 268.3 \\
      \hline
      \multirow{3}{*}{FA-SSCL} & 32-bit & 1201 &  67.2 &  26.1 &  8.5 & 207.8 & 5.1 & 345.5 \\
      %\cline{3-9}
                               & 16-bit & 1198 &  68.7 &  25.6 &  7.7 & 225.7 & 4.3 & 408.7 \\
      %\cline{3-9}
                               &  8-bit & 1259 &  71.8 &  24.4 &  7.7 & 227.3 & 4.1 & 425.9 \\
    \end{tabular}
    }}
  \end{table}

  Another important parameter in both software and hardware implementations is the quantization of data in the decoder. More specifically, the representations of LLRs and partial sums in the decoder have an impact on decoding performance. Quantized implementations of the SC algorithm have already been proposed in \cite{Giard2016} but to the best of our knowledge, the proposed decoder is the first SCL software implementation that can benefit from the 8-bit and 16-bit fixed-point representations of LLRs and internal path metrics. In the 8-bit mode LLRs and path metrics are saturated between $-127$ and $+127$ after each operation. Moreover, to avoid overflows, the path metrics are normalized after each \textit{update\_paths()} call (cf. Alg.~\ref{alg:scl}) by subtracting the smallest metric to each one of them. Figure~\ref{plot:ber_fer}a shows the BER and FER performances of the CA-SSCL decoder for 32-bit floating-point, 16-bit and 8-bit fixed-point representations. One can observe that the \texttt{REP} nodes degrade the decoding performance in a 8-bit representation because of accumulation (red triangles curve). Indeed, it is necessary to add all the LLRs of a \texttt{REP} node together in order to process it, which may lead to an overflow in the case of fixed-point representation. It can happen when the size of the repetition nodes is not limited ($\texttt{REP}_\texttt{2+}$). However, thhe size limitation of the repetition nodes to 8 ($\texttt{REP}_\texttt{8-}$) fixes this issue. In Table~\ref{tab:fixed}, maximum latency ($\mathcal{L}_{worst}$ in $\mu s$), average latency ($\mathcal{L}_{avg}$ in $\mu s$) and information throughput ($\mathcal{T}_i$ in Mb/s) are given. Note that in 8-bit configuration only the \texttt{REP}$_{\texttt{8-}}$ nodes are used. The fixed-point implementation reduces, on average, the latency. In the high SNR region, the frame errors are less frequent. Therefore, the SCL algorithm is less necessary than in low SNR regions for Adaptive SCL algorithms. As the gain of fixed-point implementation benefits more to the SC algorithm than to the SCL algorithm, the throughput is higher in high SNR regions. For instance, up to 425.9 Mb/s is achieved in 8-bit representation with the FA-SSCL decoder. Note that the improvements described in Section~\ref{sec:implem_improv} are applied to the decoders that are given in Table~\ref{tab:fixed}.

  \subsubsection{Supporting different variants of the decoding algorithms}
  
  \begin{figure}[t]
  \centering
  \includegraphics[width=0.49\textwidth]{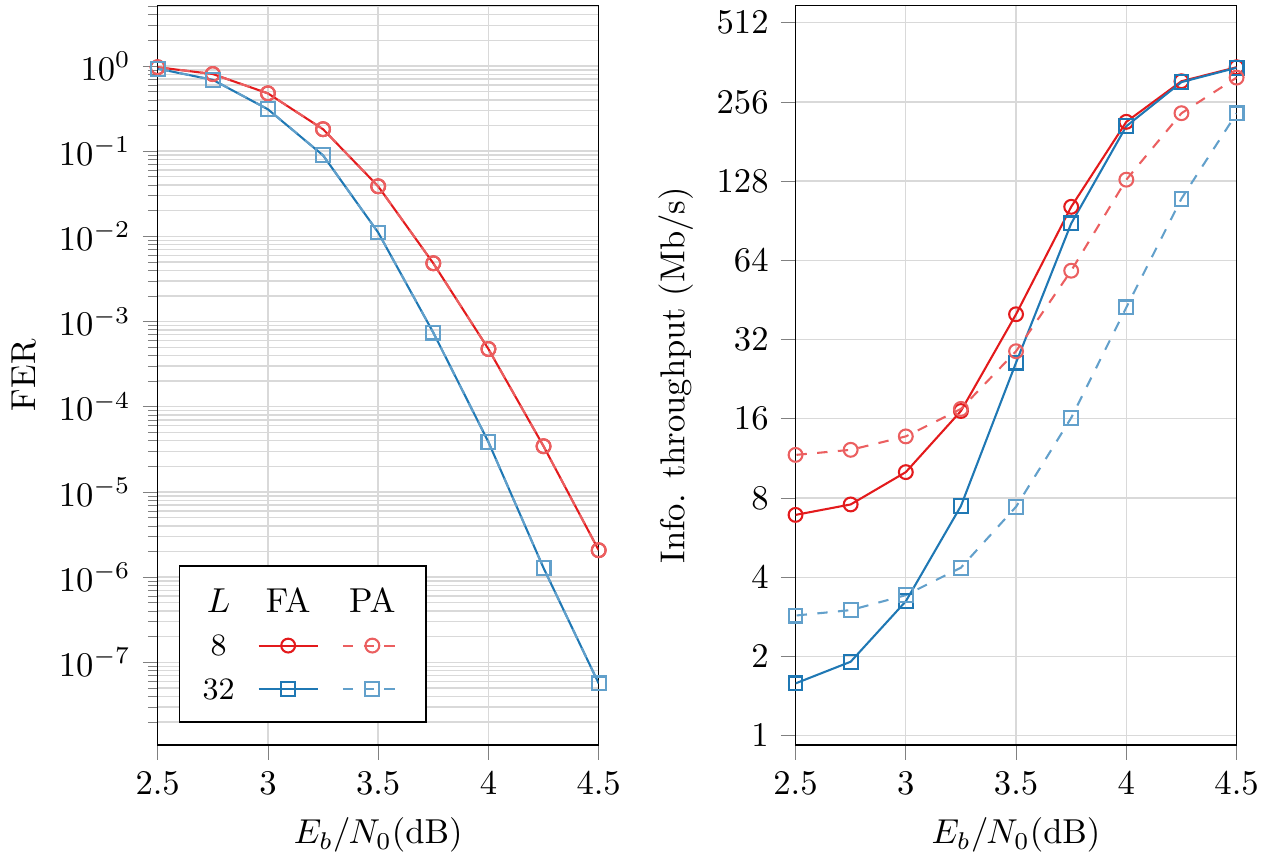}
  \caption{Frame Error Rate (FER) performance and throughput of the Fully and Partially Adaptive 
    SSCL decoders (FA and PA). Code ($2048$,$1723$) and 32-bit CRC (GZip). 32-bit floating-point representation.}
  \label{plot:ascl}
  \end{figure}

  Besides the $L$ values, the tree pruning and quantization aspects, the proposed software polar decoder supports different variants of the SCL algorithm: CA-SSCL, PA-SSCL, FA-SSCL.

  As shown in \cite{sarkis16}, the adaptive version of the SCL algorithm yields significant speedups, specially for high SNR. The original adaptive SCL described in \cite{li12}, denoted as Fully Adaptive SCL (FA-SSCL) in this paper, gradually doubles the list depth $L$ of the SCL decoder when the CRC is not valid for any of the generated codewords at a given stage until the value $L_{max}$. By contrast, the adaptive decoding algorithm implemented in \cite{sarkis16}, called in this paper Partially Adaptive SCL (PA-SSCL), directly increases the list depth from $1$ (SC) to $L_{max}$. In Figure~\ref{plot:ascl}, the two versions (FA-SSCL and PA-SSCL) are compared on a ($2048$,$1723$) polar code and 32-bit CRC (GZip). The LLRs values are based on a 32-bit floating point representation. Note that as the FER performance of PA-SSCL and FA-SSCL are exactly the same, the related error performance plots completely overlap. The throughput of the FA-SSCL algorithm is higher than that of the PA-SSCL algorithm for some SNR values, depending on the code parameters. Considering typical FER values for wireless communication standards ($10^{-3}$ to $10^{-5}$), in the case of a ($2048$,$1723$) polar code, the throughput of FA-SSCL is double that of PA-SSCL with $L = 8$, while it is multiplied by a factor of $7$ with $L=32$. The drawback of FA-SSCL is that although the average latency decreases, the worst case latency increases.

  The adaptive versions of the algorithm achieve better throughputs, but CA-SCL may also be chosen depending on the CRC. One may observe in Figure~\ref{plot:ber_fer}b that an adaptive decoder dedicated to an 8-bit CRC with a ($2048$,$1723$) polar code and $L=32$ leads to a loss of $0.5$ dB for a FER of $10^{-5}$ compared to its non adaptive counterpart.

  Both polar code genericity and decoding algorithm flexibility are helpful to support the recommendations of wireless communications in an SDR or cloud RAN context. The code and decoder parameters can be dynamically changed in the proposed decoder, while maintaining competitive throughput and latency. The following section introduces algorithmic and implementation improvements applied in the proposed decoders to keep a low decoding time.

\section{Software implementation optimizations}
\label{sec:implem_improv}
  The genericity and flexibility of the formerly described decoder prevent from using some optimizations. Unrolling the description as in \cite{sarkis16} is not possible at runtime, although code generation could be used to produce an unrolled version of any decoder as in \cite{cassagne15}. Moreover, in the case of large code lengths, the unrolling strategy can generate very large compiled binary files. This can cause instruction cache misses that would dramatically impact the decoder throughput. As this unrolling method is not applied, some implementation improvements are necessary in order to be competitive with specific decoders of the literature. The software library for polar codes from \cite{cassagne15,cassagne16_2} enables to benefit from the SIMD instructions for various target architectures. Optimizations of CRC checking benefit to both the non-adaptive and adaptive versions of the CA-SCL algorithms. The new sorting technique presented in Section~\ref{subsec:sorting} can be applied to each variation of the SCL algorithm. Finally, an efficient implementation of the partial sums memory management is proposed. It is particularly effective for short polar codes.

  \subsection{Polar Application Programming Interface}

  Reducing the decoding time with SIMD instructions is a classical technique in former software polar decoder implementations. The proposed list decoders are based on specific building blocks included from the Polar API\cite{cassagne15,cassagne16_2}.{}
  These blocks are fast and optimized implementations of the $f$, $g$, $h$ (and their variants) polar intrinsic functions. Figure~\ref{fig:f} details the SIMD implementation of these functions. This implementation is based on MIPP, a SIMD wrapper for the intrinsic functions (assembly code), and the template meta-programming technique. Consequently, the description is clear, portable, multi-format (32-bit floating-point, 16-bit and 8-bit fixed-points) and as fast as an architecture specific code.
  The \texttt{mipp::Reg<B>} and \texttt{mipp::Reg<R>} types correspond to SIMD registers. \texttt{B} and \texttt{R} define the type of the elements that are contained in this register. \texttt{B} for \textit{bit} could be \texttt{int}, \texttt{short} or \texttt{char}. \texttt{R} for \textit{real} could be \texttt{float}, \texttt{short} or \texttt{char}. In Figure~\ref{fig:f}, each operation is made on multiple elements at the same time. For instance, line 22, the addition between all the elements of the \texttt{neg\_la} and \texttt{lb} registers is executed in one CPU cycle.

  \lstset{linewidth=0.6\textwidth, xleftmargin=0.025\textwidth, xrightmargin=0.05\textwidth}

  \begin{figure}[t]
  \begin{lstlisting}[language=C++, numbers=left, numbersep=0.3em, tabsize=2, basicstyle=\footnotesize\ttfamily]
class API_polar
{
  template <typename R>
  mipp::Reg<R> f_simd(const mipp::Reg<R> &la, 
                      const mipp::Reg<R> &lb)
  {
    auto abs_la  = mipp::abs(la);
    auto abs_lb  = mipp::abs(lb);
    auto abs_min = mipp::min(abs_la, abs_lb);
    auto sign    = mipp::sign(la, lb); 
    auto lc      = mipp::neg(abs_min, sign);

    return lc;
  }

  template <typename B, typename R>
  mipp::Reg<R> g_simd(const mipp::Reg<R> &la, 
                      const mipp::Reg<R> &lb,
                      const mipp::Reg<B> &sa)
  {
    auto neg_la = mipp::neg(la, sa);
    auto lc     = neg_la + lb;

    return lc;
  }

  template <typename B>
  mipp::Reg<B> h_simd(const mipp::Reg<B>& sa,
                      const mipp::Reg<B>& sb)
  {
    return sa ^ sb;
  }
};
  \end{lstlisting}
  \caption{C++ SIMD implementation of the $f$, $g$ and $h$ functions.}
  \label{fig:f}
  \end{figure}

  In the context of software decoders, there are two well-known strategies to exploit SIMD instructions: use the elements of a register to compute 1 )many frames in parallel (INTER frame) or 2) multiple elements from a single frame (INTRA frame).
  In this paper, only the INTRA frame strategy is considered. The advantage of this strategy is the latency reduction by comparison to the INTER frame strategy.
  However, due to the nature of the polar codes, there are sometimes not enough elements to fill the SIMD registers completely. This is especially true in the nodes near the leaves.
  For this reason, SIMD instructions in the lower layers of the tree do not bring any speedup. In this context, the building blocks of the Polar API automatically switch from SIMD to sequential implementations.
  In the case of the CA-SSCL algorithm, using SIMD instructions for decoding a ($2048$, $1723$) polar code leads to an improvement of $20\%$ of the decoding throughput on average for different values of the list depth $L$.

  \subsection{Improving Cyclic Redundancy Checking}
  \label{subsec:crc_improv}

    By profiling the Adaptive SCL decoder, one may observe that a significant amount of time is spent to process the cyclic redundancy checks. Its computational complexity is O($LN$) versus the computational complexity of the SCL decoding, O($LN\log N$). The first is not negligible compared to the second.

    In the adaptive decoder, the CRC verification is performed a first time after the SC decoding. In the following, we show how to reduce the computational complexity of these CRC verifications.

    First, an efficient CRC checking code has been implemented. Whenever the decoder needs to check the CRC, the bits are packed and then computed 32 by 32. In order to further speed up the implementation, a lookup table used to store pre-computed CRC sub-sequences, and thus reduce the computational complexity. 

    After a regular SC decoding, a decision vector of size $N$ is produced. Then, the $K$ information bits must be extracted to apply cyclic redundancy check. The profiling of our decoder description shows that this extraction takes a significant amount of time compared to the check operation itself.
    Consequently, a specific extraction function was implemented. This function takes advantage of the leaf node type knowledge to perform efficient multi-element copies.

    Concerning SCL decoding, it is possible to sort the candidates according to their respective metrics and then to check the CRC of each candidate from the best to the worst. Once a candidate with a valid CRC is found, it is chosen as the decision. This method is strictly equivalent to do the cyclic redundancy check of each candidate and then to select the one with the best metric. With the adopted order, decoding time is saved by reducing the average number of checked candidates.

  \subsection{LLR and Metric Sorting}
  \label{subsec:sorting}
    Metric sorting is involved in the aforementioned path selection step, but also in the \textit{update\_paths()} sub-routine (Alg.~\ref{alg:scl}, L16) and consequently in each leaf. Sorting the LLRs is also necessary in \texttt{R1} and \texttt{SPC} nodes. Because of a lack of information about the sorting technique presented in \cite{sarkis16}, its reproduction is not possible. In the following of the paragraph the sorting algorithm used in the SCL decoder is described.

    In \texttt{R1} nodes, a Chase-$2$ \cite{chase1972class} algorithm is applied. The two maximum absolute values of the LLRs have to be identified. The way to do the minimum number of comparisons to identify the $2$ largest of $n\geq2$ elements was originally described by Schreier in \cite{schreier1932} and reported in \cite{knuth73}. The lower stages of this algorithm can be parallelized thanks to SIMD instructions in the way described in \cite{Furtak:2007:USR:1248377.1248436}. According to our experimentations, Schreier's algorithm is the most efficient compared to parallelized Batcher's merge exchange, partial quick-sort or heap-sort implemented in the C++ standard library in the case of \texttt{R1} nodes. At the end, we chose not to apply the SIMD implementation of the Schreier's algorithm because: 1) the speedup was negligible, 2) in 8-bit fixed-point, only $N \leq 256$ codewords can be considered.

    Concerning path metrics, partial quick-sort appeared to yield no gains in terms of throughput by comparison with the algorithm in \cite{schreier1932}, neither did heap-sort or parallelized Batcher's merge exchange. For a matter of consistency, only Schreier's algorithm is used in the proposed decoder, for both LLR sorting in \texttt{R1} and \texttt{SPC} nodes and for path metrics sorting. The sorting of path metrics is applied to choose the paths to be removed, kept or duplicated.

    \subsection{Partial Sum Memory Management}

    \begin{figure}[t]
    \centering
    \includegraphics[width=0.49\textwidth]{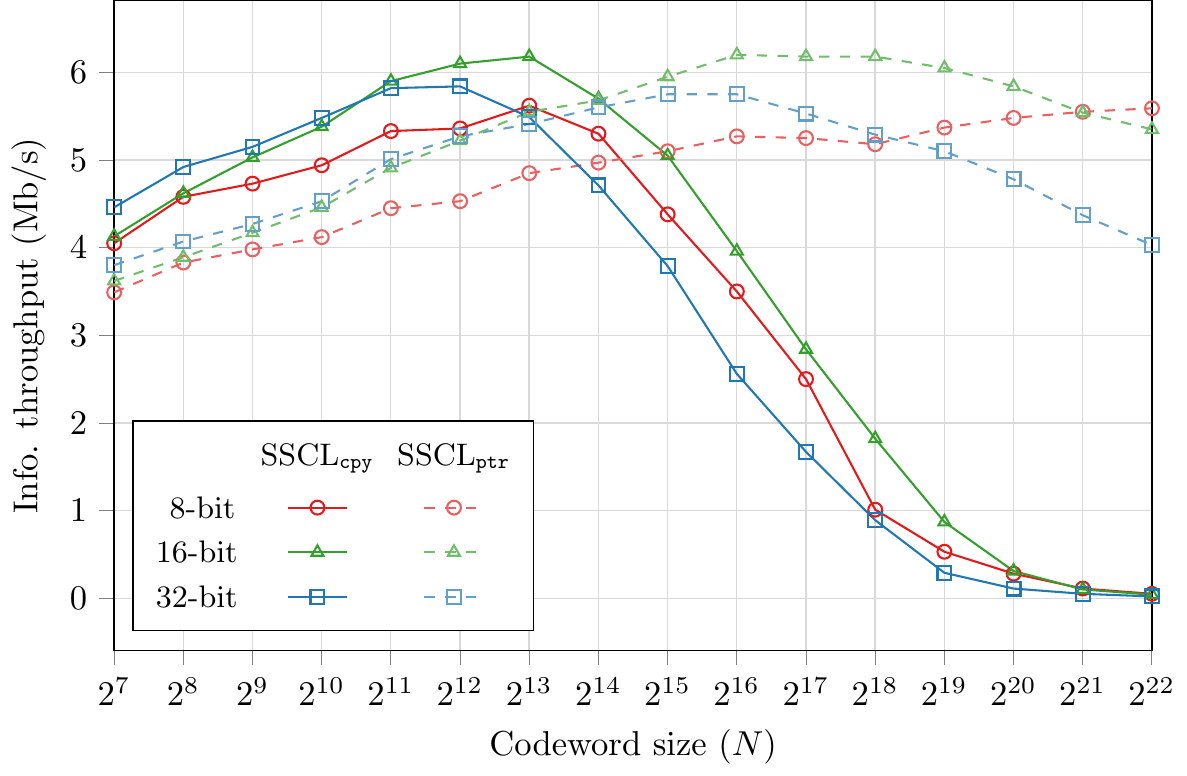}
    \caption{Information throughput of the SSCL decoder depending on the codeword
      size ($N$) and the partial sums management. $R = 1 / 2$, $L = 8$.}
    \label{plot:thr}
    \end{figure}

    An SCL decoder can be seen as $L$ replications of an SC decoder. The first possible memory layout is the one given in Figure~\ref{fig:dec_tree}. In this layout, the partial sums $\hat{s}$ of each node is stored in a dedicated array. Therefore, a memory of size $2N-1$ bits is necessary in the SC decoder, or $L \times (2N -1)$ bits in the SCL decoder. This memory layout is described in \cite{tal12} and applied in previous software implementations \cite{sarkis14_3,sarkis16,shen16}.

    A possible improvement is to change the memory layout to reduce its footprint. Due to the order of operations in both SC and SCL algorithms, the partial sums on a given layer are only used once by the $\bm{h}$ function and can then be overwritten. Thus, a dedicated memory allocation is not necessary at each layer of the tree. The memory can be shared between the stages. Therefore the memory footprint can be reduced from $2N-1$ to $N$ in the SC decoder as shown in \cite{leroux2013semi}. A reduction from $L \times (2N -1)$ to $LN$ can be obtained in the SCL decoder.

    In the case of the SCL algorithm, $L$ paths have to be assigned to $L$ partial sum memory arrays. In \cite{tal12}, this assignment is made with pointers. The advantage of pointers is that when a path is duplicated, in the \textit{update\_paths()} sub-routine of Alg.~\ref{alg:scl}, the partial sums are not copied. Actually, they can be shared between paths thanks to the use of pointers. This method limits the number of memory transactions. Unfortunately, it is not possible to take advantage of the memory space reduction: the partial sums have to be stored on $L \times (2N -1)$ bits. There is an alternative to this mechanism. If a logical path is statically assigned to a memory array, no pointers are necessary at the cost that partial sums must be copied when a path is duplicated (only $LN$ bits are required). This method is called SSCL$_{\texttt{cpy}}$ whereas the former is called SSCL$_{\texttt{ptr}}$.

    Our experiments have proved that the overhead of handling pointers plus the extra memory space requirement cause the SSCL$_{\texttt{cpy}}$ to be more efficient than the SSCL$_{\texttt{ptr}}$ for short and medium code lengths, as shown in Figure~\ref{plot:thr}. The 32-bit version uses floating-point LLRs, whereas 16-bit and 8-bit versions are in fixed-point.
    Notice that in this work, each bit of the partial sums is stored on an 8-bit, 16-bit or 32-bit number accordingly to the LLR data type.
    The code rate $R$ is equal to $1/2$. The throughput of the SSCL$_{\texttt{cpy}}$ version is higher for $N \leq 8192$ whereas the SSCL$_{\texttt{ptr}}$ version is more efficient for higher values of $N$. Although it does not appear in  Figure~\ref{plot:thr}, experiments showed that the lower $L$ is, the more efficient SSCL$_{\texttt{cpy}}$ is compared to SSCL$_{\texttt{ptr}}$. Figure~\ref{plot:thr} also illustrates the impact of the representation of partial sums. For very high values of $N$, 8-bit fixed point representation takes advantage of fewer cache misses. According to the results presented in Figure~\ref{plot:code}, as the decoding performance improvements of the SCL algorithm are not very significant compared to the SC algorithm for long polar codes, SSCL$_{\texttt{cpy}}$ is the appropriate solution in most practical cases. 

    In our decoder description, LLRs are managed with pointers, as it is the case in other software implementations of the literature \cite{sarkis14_3,sarkis16,shen16}. We tried to remove the pointer handling as for the partial sums, but it appeared that it was not beneficial in any use case.

\section{Experiments and Measurements}
\label{sec:measures}

  Throughput and latency measurements are detailed in this section. The proposed decoder implementation is compared with the previous software decoders. Despite the additional levels of genericity and flexibility, the proposed implementation is very competitive with its counterparts. Note that all the results presented in the following can be reproduced with the AFF3CT tool.

  During our investigations, all the throughput and latency measurements have been obtained on a single core of an Intel i5-6600K CPU (Skylake architecture with AVX2 SIMD) with a base clock frequency of 3.6 GHz and a maximum turbo frequency of 3.9 GHz.
  The description has been compiled on Linux with the C++ GNU compiler (version 5.4.0) and with the following options: \texttt{-Ofast -march=native -funroll-loops}.
  \subsection{Fully Adaptive SCL}

    Being able to easily change the list size of the SCL decoders enables the use of the FA-SSCL algorithm. With an unrolled decoder as proposed in \cite{sarkis16}, the fully adaptive decoder would imply to generate a fully unrolled decoder for each value of the list depth. In our work, only one source code gives the designer the possibility to run each variation of the SCL decoders. FA-SSCL algorithm is the key to achieve the highest possible throughput. As shown in Table~\ref{tab:fixed}, with an 8-bit fixed point representation of the decoder inner values, the achieved throughput in the case of the ($2048$,$1723$) polar code is about $425$ Mb/s on the i5-6600K for an $E_b/N_0$ value of $4.5$ dB. It corresponds to a FER of $5\times10^{-8}$. This throughput is almost 2 times higher than the throughput of the PA-SSCL algorithm. The highest throughput increase from PA-SSCL to FA-SSCL, of about $380\%$, is in the domain where the FER is between $10^{-3}$ and $10^{-5}$. It is the targeted domain for wireless communications like LTE or 5G. In these conditions, the throughput of FA-SSCL algorithm is about $227$ Mb/s compared to $42$ Mb/s for the PA-SSCL algorithm.

    In Adaptive SCL algorithms, the worst case latency is the sum of the latency of each triggered algorithm. In the case of PA-SSCL with $L_{max}=32$, it is just the sum of the latency of the SC algorithm, plus the latency of the SCL algorithm with $L=32$. In the case of the FA-SSCL algorithm, it is the sum of the decoding latency of the SC algorithm and all the decoding latencies of the SCL algorithm for $L={2,4,8,16,32}$. This is the reason why the worst latency of the PA-SSCL algorithm is lower while the average latency and consequently the average throughput is better with the FA-SSCL algorithm.

    \begin{table}[t]
      \centering
      \caption{Throughput and latency comparison with state-of-the-art SCL decoders. 32-bit floating-point representation.
      \newline
      Code (2048,1723), $L = 32$, 32-bit CRC.}
      \label{tab:res}
      {\small\resizebox{\linewidth}{!}{ 
      \begin{tabular}{r|r|c|c c c}
        \multirow{2}{*}{\textbf{Target}} & \multirow{2}{*}{\textbf{Decoder}}  & \multirow{1}{*}{\textbf{$\bm{\mathcal{L}_{worst}}$}} & \multicolumn{3}{c}{$\bm{\mathcal{T}_i}$ (Mb/s)} \\
        \cline{4-6}
        &                           & ($\mu s$)                          & \textbf{3.5 dB} & \textbf{4.0 dB} & \textbf{4.5 dB} \\
        \hline
        % \hline
        \multirow{1}{*}{i7-4790K}
        & CA-SCL \cite{shen16    } & 1572                           & 1.10            & 1.10            & 1.10            \\
        \hline
        \multirow{3}{*}{i7-2600}
        & CA-SCL \cite{sarkis14_3} & 23000                          & 0.07            & 0.07            & 0.07            \\
        & CA-SSCL\cite{sarkis14_3} & 3300                           & 0.52            & 0.52            & 0.52            \\
        & PA-SSCL \cite{sarkis14_3} & $\approx$ 3300                 & 0.9             & 4.90            & 54.0            \\
        \hline
        \multirow{3}{*}{i7-2600}
        & CA-SCL \cite{sarkis16}   & 2294                           & 0.76            & 0.76            & 0.76            \\
        & CA-SSCL\cite{sarkis16}   & 433                            & 4.0             & 4.0             & 4.0             \\
        & PA-SSCL \cite{sarkis16}   & $\approx$ 433                  & 8.6             & 33.0            & 196.0           \\
        \hline
  %     original data
  %     \multirow{4}{*}{\rotatebox[origin=c]{90}{\textbf{E5-2650}}}
  %     & Proposed CA-SCL          & 6554                           & 0.27            & 0.27            & 0.27            \\
  %     & Proposed CA-SSCL         & 1048                           & 1.67            & 1.67            & 1.67            \\
  %     & Proposed PA-SSCL          & $\approx$ 1048                 & 4.07            & 22.9            & 124.1           \\
  %     & Proposed FA-SSCL          & $\approx$ 2096                 & 14.3            & 109.8           & 180.0           \\
  %     \hline
  %     rescaled data from E5-2650
        \multirow{4}{*}{i7-2600}
        & This CA-SCL              & 4819                           & 0.37            & 0.37            & 0.37            \\
        & This CA-SSCL             & 770                            & 2.3             & 2.3             & 2.3             \\
        & This PA-SSCL              & 847                            & 5.5             & 31.1            & 168.4           \\
        & This FA-SSCL              & 1602                           & 19.4            & 149.0           & 244.3           \\
        \hline
        \multirow{4}{*}{i5-6600K}
        & This CA-SCL              & 3635                           & 0.48            & 0.48            & 0.48            \\
        & This CA-SSCL             & 577                            & 3.0             & 3.0             & 3.0             \\
        & This PA-SSCL              & 635                            & 7.6             & 42.1            & 237.6           \\
        & This FA-SSCL              & 1201                           & 26.1            & 207.8           & 345.5           \\
      \end{tabular}
      }}
    \end{table}

  \subsection{Comparison With State-Of-The-Art SCL Decoders.}

    The throughput and latency of the proposed decoder compared to other reported implementations are detailed in Table~\ref{tab:res}. For all the decoders, all the available tree pruning optimizations are applied excluding the \texttt{SPC4+} nodes because of the performance degradation. Each decoder is based on a 32-bit floating-point representation. The polar code parameters are $N=2048$, $K=1723$ and the 32-bit GZip CRC is used. The list size is $L=32$.

    The latency given in Table~\ref{tab:res} is the worst case latency and the throughput is the average information throughput. The first version, CA-SCL, is the implementation of the CA-SCL algorithm without any tree pruning. As mentioned before the throughput of the proposed CA-SSCL decoder ($2.3$ Mb/s) is only halved compared to the specific unrolled CA-SSCL decoder described in \cite{sarkis16} (4.0 Mb/s). The proposed CA-SSCL decoder is approximately 4 times faster than the generic implementation in \cite{sarkis14_3} ($0.52$ Mb/s) and 2 times faster than the CA-SCL implementation in \cite{shen16} ($1.1$ Mb/s) thanks to the implementation improvements detailed in Section \ref{sec:implem_improv}.
    Furthermore, the proposed decoder exhibits a much deeper level of genericity and flexibility than the ones proposed in \cite{sarkis14_3,shen16}. Indeed, the following features were not enabled: the customization of the tree pruning, the 8-bit and 16-bit fixed-point representations of the LLRs, the puncturing patterns and the FA-SSCL algorithm.

    When implemented on the same target (i7-2600), the proposed PA-SSCL is competitive with the unrolled PA-SSCL in \cite{sarkis16}, being only two times slower.
    This can be explained by the improvements concerning the CRC that are described in Section \ref{subsec:crc_improv}, especially the information bits extraction in the SC decoder. Finally, as mentioned before, the throughput of the proposed FA-SSCL significantly outperforms all the other SCL decoders (up to 345.5 Mb/s at 4.5 dB in 32-bit floating-point).

\section{Conclusion}
\label{sec:conc}

The trend towards Cloud RAN networks in the context of mobile communications and the upcoming 5G standard motivated an investigation of  the possibility of implementing generic and flexible software polar decoders. Means of implementing such flexible decoders are reported in this paper. A single source code is necessary to address any code lengths, code rates, frozen bits sets, puncturing patterns and cyclic redundancy check polynomials. 

This genericity is obtained without sacrificing the throughput of the decoders, thanks to the possibility to adjust the decoding algorithm and the possibility to apply multiple implementation related and algorithmic optimizations. In fact, to the best of our knowledge, the proposed adaptive SCL decoder is the fastest to be found in the literature, with a throughput of 425 Mb/s on a single core for $N = 2048$ and $K = 1723$ at 4.5 dB.

Being included in the open-source AFF3CT tool, all the results presented in this paper can be easily reproduced. Moreover, this tool can be used for polar codes exploration, which is of interest for the definition of digital communication standards and for practical implementations in an SDR environment.

\section*{Acknowledgments}
The authors would like to thank the Natural Sciences and Engineering Research Council of Canada, Prompt, and Huawei Technologies Canada Co. Ltd. for financial support to this project.
This work was also supported by a grant overseen by the French National Research Agency (ANR), ANR-15-CE25-0006-01.

\bibliographystyle{IEEEtran}
\bibliography{article}

\begin{thebibliography}{10}
\providecommand{\url}[1]{#1}
\csname url@samestyle\endcsname
\providecommand{\newblock}{\relax}
\providecommand{\bibinfo}[2]{#2}
\providecommand{\BIBentrySTDinterwordspacing}{\spaceskip=0pt\relax}
\providecommand{\BIBentryALTinterwordstretchfactor}{4}
\providecommand{\BIBentryALTinterwordspacing}{\spaceskip=\fontdimen2\font plus
\BIBentryALTinterwordstretchfactor\fontdimen3\font minus
  \fontdimen4\font\relax}
\providecommand{\BIBforeignlanguage}[2]{{%
\expandafter\ifx\csname l@#1\endcsname\relax
\typeout{** WARNING: IEEEtran.bst: No hyphenation pattern has been}%
\typeout{** loaded for the language `#1'. Using the pattern for}%
\typeout{** the default language instead.}%
\else
\language=\csname l@#1\endcsname
\fi
#2}}
\providecommand{\BIBdecl}{\relax}
\BIBdecl

\bibitem{arikan09}
E.~Arikan, ``Channel polarization: a method for constructing capacity-achieving
  codes for symmetric binary-input memoryless channels,'' \emph{IEEE
  Transactions on Information Theory (TIT)}, vol.~55, no.~7, pp. 3051--3073,
  2009.

\bibitem{tal12}
I.~Tal and A.~Vardy, ``List decoding of polar codes,'' in \emph{Proceedings of
  the IEEE International Symposium on Information Theory (ISIT)}, 2011, pp.
  1--5.

\bibitem{3GPP_16}
``{3GPP} {TSG} {RAN} {WG1} {meeting} \#87, {Chairman’s} {notes} of {agenda}
  {item} 7.1.5 {Channel} coding and modulation,'' 2016.

\bibitem{wubben2014benefits}
D.~W{\"u}bben, P.~Rost, J.~S. Bartelt, M.~Lalam, V.~Savin, M.~Gorgoglione,
  A.~Dekorsy, and G.~Fettweis, ``Benefits and impact of cloud computing on {5G}
  signal processing: flexible centralization through cloud-ran,'' \emph{IEEE
  Signal Processing Magazine}, vol.~31, no.~6, pp. 35--44, 2014.

\bibitem{rost2014cloud}
P.~Rost, C.~J. Bernardos, A.~De~Domenico, M.~Di~Girolamo, M.~Lalam, A.~Maeder,
  D.~Sabella, and D.~W{\"u}bben, ``Cloud technologies for flexible {5G} radio
  access networks,'' \emph{IEEE Communications Magazine}, vol.~52, no.~5, pp.
  68--76, 2014.

\bibitem{ericsson-wp-cloud-ran}
\BIBentryALTinterwordspacing
Ericsson, ``Cloud ran - the benefits of cirtualization, centralisation and
  coordination,'' Tech. Rep., 2015. [Online]. Available:
  \url{https://www.ericsson.com/assets/local/publications/white-papers/wp-cloud-ran.pdf}
\BIBentrySTDinterwordspacing

\bibitem{huawei-5G}
\BIBentryALTinterwordspacing
Huawei, ``{5G}: A technology vision,'' Tech. Rep., 2013. [Online]. Available:
  \url{https://www.huawei.com/ilink/en/download/HW_314849}
\BIBentrySTDinterwordspacing

\bibitem{rodriguez2017towards}
V.~Q. Rodriguez and F.~Guillemin, ``Towards the deployment of a fully
  centralized cloud-ran architecture,'' in \emph{Proceedings of the IEEE
  International Wireless Communications and Mobile Computing Conference
  (IWCMC)}, 2017, pp. 1055--1060.

\bibitem{nikaein2015processing}
N.~Nikaein, ``Processing radio access network functions in the cloud: critical
  issues and modeling,'' in \emph{Proceedings of the ACM International Workshop
  on Mobile Cloud Computing and Services (MCS)}, 2015, pp. 36--43.

\bibitem{sarkis16}
G.~Sarkis, P.~Giard, A.~Vardy, C.~Thibeault, and W.~J. Gross, ``Fast list
  decoders for polar codes,'' \emph{IEEE Journal on Selected Areas in
  Communications (JSAC)}, vol.~34, no.~2, pp. 318--328, 2016.

\bibitem{sarkis14_3}
------, ``Increasing the speed of polar list decoders,'' in \emph{Proceedings
  of the IEEE International Workshop on Signal Processing Systems (SiPS)},
  2014, pp. 1--6.

\bibitem{cassagne15}
A.~Cassagne, B.~{Le Gal}, C.~Leroux, O.~Aumage, and D.~Barthou, ``An efficient,
  portable and generic library for successive cancellation decoding of polar
  codes,'' in \emph{Proceedings of the Springer International Workshop on
  Languages and Compilers for Parallel Computing (LCPC)}, 2015, pp. 303--317.

\bibitem{cassagne16_2}
A.~Cassagne, O.~Aumage, C.~Leroux, D.~Barthou, and B.~Le~Gal, ``Energy
  consumption analysis of software polar decoders on low power processors,'' in
  \emph{Proceedings of the IEEE European Signal Processing Conference
  (EUSIPCO)}, 2016, pp. 642--646.

\bibitem{6557004}
I.~Tal and A.~Vardy, ``How to construct polar codes,'' \emph{IEEE Transactions
  on Information Theory (TIT)}, vol.~59, no.~10, pp. 6562--6582, Oct 2013.

\bibitem{6279525}
P.~Trifonov, ``Efficient design and decoding of polar codes,'' \emph{IEEE
  Transactions on Communications}, vol.~60, no.~11, pp. 3221--3227, November
  2012.

\bibitem{legal15}
B.~{Le Gal}, C.~Leroux, and C.~J\'ego, ``Multi-{G}b/s software decoding of
  polar codes,'' \emph{IEEE Transactions on Signal Processing (TSP)}, vol.~63,
  no.~2, pp. 349--359, 2015.

\bibitem{sarkis14_1}
G.~Sarkis, P.~Giard, A.~Vardy, C.~Thibeault, and W.~J. Gross, ``Fast polar
  decoders: algorithm and implementation,'' \emph{IEEE Journal on Selected
  Areas in Communications (JSAC)}, vol.~32, no.~5, pp. 946--957, 2014.

\bibitem{balatsoukas2015llr}
A.~Balatsoukas-Stimming, M.~B. Parizi, and A.~Burg, ``{LLR}-based successive
  cancellation list decoding of polar codes,'' \emph{IEEE Transactions on
  Signal Processing (TSP)}, vol.~63, no.~19, pp. 5165--5179, 2015.

\bibitem{alamdar-yazdi11}
A.~Alamdar-Yazdi and F.~Kschischang, ``A simplified successive-cancellation
  decoder for polar codes,'' \emph{IEEE Communications Letters}, vol.~15,
  no.~12, pp. 1378--1380, 2011.

\bibitem{li12}
B.~Li, H.~Shen, and D.~Tse, ``An adaptive successive cancellation list decoder
  for polar codes with cyclic redundancy check,'' \emph{IEEE Communications
  Letters}, vol.~16, no.~12, pp. 2044--2047, December 2012.

\bibitem{matsumoto1998mersenne}
M.~Matsumoto and T.~Nishimura, ``Mersenne twister: a 623-dimensionally
  equidistributed uniform pseudo-random number generator,'' \emph{ACM
  Transactions on Modeling and Computer Simulation (TOMACS)}, vol.~8, no.~1,
  pp. 3--30, 1998.

\bibitem{box1958note}
G.~E.~P. Box, M.~E. Muller \emph{et~al.}, ``A note on the generation of random
  normal deviates,'' \emph{The annals of mathematical statistics}, vol.~29,
  no.~2, pp. 610--611, 1958.

\bibitem{dahlman20134g}
E.~Dahlman, S.~Parkvall, and J.~Skold, \emph{4G: LTE/LTE-advanced for mobile
  broadband}.\hskip 1em plus 0.5em minus 0.4em\relax Academic press, 2013.

\bibitem{6936302}
R.~Wang and R.~Liu, ``A novel puncturing scheme for polar codes,'' \emph{IEEE
  Communications Letters}, vol.~18, no.~12, pp. 2081--2084, Dec 2014.

\bibitem{6655078}
K.~Niu, K.~Chen, and J.~R. Lin, ``Beyond turbo codes: rate-compatible punctured
  polar codes,'' in \emph{Proceedings of the IEEE International Conference on
  Communications (ICC)}, June 2013, pp. 3423--3427.

\bibitem{7152894}
V.~Miloslavskaya, ``Shortened polar codes,'' \emph{IEEE Transactions on
  Information Theory (TIT)}, vol.~61, no.~9, pp. 4852--4865, Sept 2015.

\bibitem{CRCWiki}
``Cyclic redundancy check,''
  \url{https://en.wikipedia.org/wiki/Cyclic_redundancy_check}, accessed:
  2017-03-13.

\bibitem{zhang2017crc}
Q.~Zhang, A.~Liu, X.~Pan, and K.~Pan, ``{CRC} code design for list decoding of
  polar codes,'' \emph{IEEE Communications Letters}, vol.~21, no.~6, pp.
  1229--1232, 2017.

\bibitem{lin2014reduced}
J.~Lin, C.~Xiong, and Z.~Yan, ``A reduced latency list decoding algorithm for
  polar codes,'' in \emph{Proceedings of the IEEE International Workshop on
  Signal Processing Systems (SiPS)}, 2014, pp. 1--6.

\bibitem{Giard2016}
P.~Giard, G.~Sarkis, C.~Leroux, C.~Thibeault, and W.~J. Gross, ``Low-latency
  software polar decoders,'' \emph{Springer Journal of Signal Processing
  Systems (JSPS)}, pp. 31--53, Jul 2016.

\bibitem{chase1972class}
D.~Chase, ``Class of algorithms for decoding block codes with channel
  measurement information,'' \emph{IEEE Transactions on Information Theory
  (TIT)}, vol.~18, no.~1, pp. 170--182, 1972.

\bibitem{schreier1932}
J.~Schreier, ``On tournament elimination systems,'' \emph{Mathesis Polska},
  vol.~7, pp. 154--160, 1932.

\bibitem{knuth73}
D.~Knuth, \emph{The art of computer programming}.\hskip 1em plus 0.5em minus
  0.4em\relax Addison-Wesley, 1973, no.~3.

\bibitem{Furtak:2007:USR:1248377.1248436}
T.~Furtak, J.~N. Amaral, and R.~Niewiadomski, ``Using {SIMD} registers and
  instructions to enable instruction-level parallelism in sorting algorithms,''
  in \emph{Proceedings of the ACM Symposium on Parallel Algorithms and
  Architectures}, 2007, pp. 348--357.

\bibitem{shen16}
Y.~Shen, C.~Zhang, J.~Yang, S.~Zhang, and X.~You, ``Low-latency software
  successive cancellation list polar decoder using stage-located copy,'' in
  \emph{Proceedings of the IEEE International Conference on Digital Signal
  Processing (DSP)}, 2016.

\bibitem{leroux2013semi}
C.~Leroux, A.~J. Raymond, G.~Sarkis, and W.~J. Gross, ``A semi-parallel
  successive-cancellation decoder for polar codes,'' \emph{IEEE Transactions on
  Signal Processing (TSP)}, vol.~61, no.~2, pp. 289--299, 2013.

\end{thebibliography}

\end{document}